\documentclass[twoside,english,5p]{elsarticle}
\usepackage[T1]{fontenc}
\usepackage[latin9]{inputenc}
\usepackage{array}
\usepackage{float}
\usepackage{multirow}
\usepackage{amsmath}
\usepackage{graphicx}

\makeatletter

\providecommand{\tabularnewline}{\\}

\journal{ArXiv}

\makeatother

\usepackage{babel}
\begin{document}

\begin{frontmatter}{}

\title{\textbf{\Large{}Sensor Defense In-Software (SDI):}\\
\textbf{\Large{}Practical Software Based Detection of Spoofing Attacks
on Position Sensors }}

\author[bgu]{Kevin Sam Tharayil\corref{cor1}}

\ead{tharayil@post.bgu.ac.il}

\author[bgu]{Benyamin Farshteindiker}

\ead{farshben@post.bgu.ac.il}

\author[bgu]{Shaked Eyal}

\ead{shakedey@post.bgu.ac.il}

\author[bgu]{Nir Hasidim}

\ead{hasidimn@post.bgu.ac.il}

\author[bgu]{Roy Hershkovitz}

\ead{royhersh@post.bgu.ac.il}

\author[bgu]{Shani Houri}

\ead{hourish@post.bgu.ac.il}

\author[bgu]{Ilia Yoffe (Iofedov)}

\ead{iofedov@post.bgu.ac.il}

\author[tomer]{Michal Oren}

\ead{michaloren78@gmail.com}

\author[bgu]{Yossi Oren}

\ead{yos@bgu.ac.il}

\cortext[cor1]{Corresponding author}

\address[bgu]{Department of Software and Information Systems Engineering, Ben Gurion
University, Israel}

\address[tomer]{Tomer Ltd., Israel}
\begin{abstract}
Position sensors, such as the gyroscope, the magnetometer and the
accelerometer, are found in a staggering variety of devices, from
smartphones and UAVs to autonomous robots. Several works have shown
how adversaries can mount spoofing attacks to remotely corrupt or
even completely control the outputs of these sensors. With more and
more critical applications relying on sensor readings to make important
decisions, defending sensors from these attacks is of prime importance. 

In this work we present practical software based defenses against
attacks on two common types of position sensors, specifically the
gyroscope and the magnetometer. We first characterize the sensitivity
of these sensors to acoustic and magnetic adversaries. Next, we present
two software-only defenses: a machine learning based single sensor
defense, and a sensor fusion defense which makes use of the mathematical
relationship between the two sensors. We performed a detailed theoretical
analysis of our defenses, and implemented them on a variety of smartphones,
as well as on a resource-constrained IoT sensor node. Our defenses
do not require any hardware or OS-level modifications, making it possible
to use them with existing hardware. Moreover, they provide a high
detection accuracy, a short detection time and a reasonable power
consumption.
\end{abstract}

\end{frontmatter}{}

\section{Introduction}

Many electronic devices, such as smartphones and sensor nodes, are
equipped with position sensors. These sensors are capable of measuring
the position, orientation and motion of the device in three-dimensional
space. We rely on these sensors for increasingly sensitive tasks including
authentication\citet{DBLP:conf/ccs/ContiZC11,DBLP:conf/isca/LeeL16},
navigation\citet{DBLP:conf/huc/LiZDGLZ12}, and health monitoring\citet{10.1371/journal.pone.0141694}.
This paper focuses on two widely used sensors: the gyroscope, which
measures a device's angular momentum, or rate of rotation, and the
magnetometer, which measures a device's orientation with respect to
the magnetic field of the Earth.

\begin{figure}
\begin{centering}
\includegraphics[width=1.1\columnwidth]{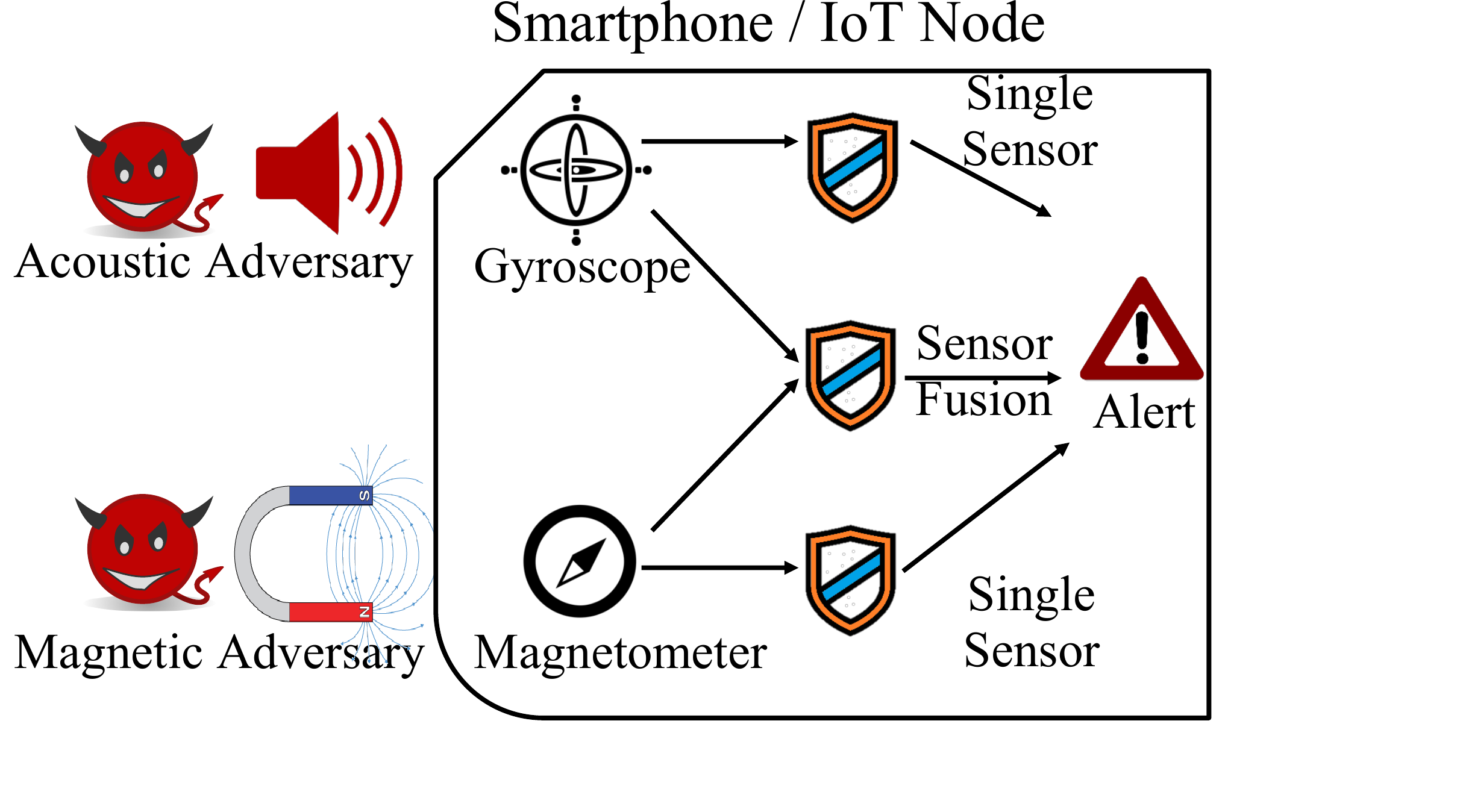}
\par\end{centering}
\caption{\label{fig:Overall-Description}Overall Description of our Defenses}
\end{figure}

Several recent works have shown how the readings of these sensors
can be spoofed by applying an external acoustic stimulus to the device
or its surroundings \citet{walnut,DBLP:conf/uss/TuLLH18}. The spoofed
output of a sensor does not reflect the device\textquoteright s actual
rotation or orientation; instead, the output is overwritten by artificial
values which are either randomly corrupted or completely controlled
by the attacker. Sensor spoofing attacks on smartphones are already
being used for malicious purposes. For example, the online publication
\textit{Sixth Tone} reported on June 2018 that Chinese university
students, who are required to reach at least 10,000 steps per day
as part of their fitness requirement, use a variety of devices called
``WeRun Boosters'' to spoof the motion sensors on their smartphones,
generating 6,000 to 7,000 steps on a smartphone per hour\citet{Stepcounter}.
The risks associated with sensor spoofing will only grow as the amount
of sensitive applications relying on these sensors increases. For
example, Wang et al. \citet{wang} and Reinertsen et al.\citet{0967-3334-38-7-1456}
proposed to use sensor measurements to assess the severity of illness
of patients with schizophrenia. Sensor spoofing attacks, when applied
to this scenario, may erroneously cause a person to be hospitalized
in a psychiatric ward. 

While several papers have discussed sensor spoofing, few of them have
discussed the prevention of these attacks, a gap we wish to address
in this work. One of the main limitations of many defenses against
sensor spoofing is that they either require changes to the sensor
hardware or to the low-level firmware used to interface it to the
phone's CPU. Since position sensors are typically highly integrated
low cost devices with a relatively long development cycles, such modifications
are difficult to apply to hardware already deployed in the field and,
are hard to justify from a system integration standpoint. While software-based
anomaly detection mechanisms have been proposed for other types of
sensor systems, such as wireless sensor networks \citet{DBLP:conf/aina/PintoLPPS18},
they typically did not consider a malicious adversary but only a random
fault model.

\textbf{Our Contribution: }In this paper we propose two software-based
defense methods against acoustic and magnetic attacks on a phone's
gyroscope and magnetometer. Our first defense method, SDI-1, uses
machine learning to detect anomalies in the output of a single sensor.
This defense method can detect sensor corruption attacks, but cannot
detect cases where a more powerful adversary can force the sensor
to output a spoofed but valid reading. Our second defense method,
SDI-2, applies sensor fusion to compare the readings of multiple sensors
measuring a similar type of motion. This method can potentially protect
against a more powerful sensor spoofing adversary, as long as this
adversary cannot control the entire set of sensors available on the
device. Specifically, in this paper we present single-sensor defenses
for acoustic attacks on the gyroscope and for magnetic attacks on
the magnetometer. We also present a sensor fusion based defense combining
the gyroscope and the magnetometer. We describe the physical and mathematical
relationship between expected sensor readings, and show how the defender
can measure deviations between the two sensors to detect an attack.
We implemented our defenses on multiple smartphones from different
vendors, as well as on a resource-constrained IoT node, in each case
measuring the accuracy, detection time and power usage of our defenses.
The main advantage of these defenses are that they are purely software
based, and can therefore be deployed on many types of devices without
any hardware modification.

\textbf{Document Structure:} We begin by describing the spoofing attacks
on the MEMS gyroscope and magnetometer. In Section \ref{sec:Defense-Methods}
we describe SDI-1, a machine learning-based single sensor defense,
and SDI-2, a sensor fusion-based single sensor defense, and show how
they can protect against acoustic and magnetic attacks on the gyroscope
and on the magnetometer respectively. In Section \ref{sec:Evaluation}
we perform a practical evaluation of our defense methods. Finally,
in Section \ref{sec:Discussion} we discuss defenses for another type
of sensor, the accelerometer, and conclude by discussing further applications
of sensor fusion and its improvements. 

\subsection{Types of Position Sensors}

A smartphone\textquoteright s various position sensors are used to
measure the phone\textquoteright s position and motion in space along
the six axes of motion (or six degrees of freedom). The measurements
of the device\textquoteright s sensors are generally provided in the
device\textquoteright s frame of reference: a Cartesian coordinate
system with coordinates attached to the device. This coordinate system
is rotated with respect to the world\textquoteright s frame of reference,
which is a standard static coordinate system. Of the six degrees of
freedom, three coordinates (X, Y, and Z) are used to describe the
phone\textquoteright s position and linear motion in space, while
the three other coordinates ($\rho$,$\phi$ and $\theta$, or pitch,
roll and yaw) are used to describe the phone\textquoteright s Cartesian
axes orientation with respect to the world\textquoteright s frame
of reference and its rotational motion.

The \textbf{gyroscope} is a MEMS-based sensor which measures the device\textquoteright s
angular velocity in units of radian per second. As described in \citet{rockingdrones},
Microelectromechanical systems (MEMS) gyroscopes typically contain
a small mass moving back and forth at a constant frequency. As the
phone is rotated, the Coriolis effect acts on this moving mass and
causes it to vibrate with an amplitude that is directly related to
the angular rotation rate. The modulated vibration amplitude is then
converted to voltage, typically by a capacitive or piezo-electric
sensor. The \textbf{magnetometer}, or compass, measures the direction
and magnitude of the ambient magnetic field around the device, in
units of microtesla. As described in\citet{jiang}, virtually all
smartphones use a Hall effect magnetometer, which works by detecting
the voltage differential induced by the Hall effect across a thin
metallic surface in response to a magnetic field perpendicular to
the surface. The magnetic field measured by the phone field is typically
a combination of the Earth\textquoteright s magnetic field, which
points more or less to the north, and additional magnetic sources
in the vicinity of the phone, such as iron beams, electric motors
or induction coils. As long as the phone stays in the same place and
the additional magnetic sources stay constant over time, the magnetometer\textquoteright s
reading will point to the same direction in the world\textquoteright s
reference frame, even when the phone is rotated. Other common position
sensors include the accelerometer, which measures the linear acceleration
of the device and the GPS sensor which measures the location of the
device on Earth.

\subsection{Spoofing Attacks on Position Sensors }

As mentioned in the previous section, MEMS gyroscopes contain a small
moving mass. As shown in \citet{DBLP:conf/uss/TuLLH18} and \citet{rockingdrones},
they are vulnerable to acoustic attacks, in which the sensor is subjected
to external vibrations with the sensor's mechanical resonant frequency.
When the moving mass inside the sensor is stimulated by this acoustic
signal, it begins vibrating with a high amplitude. This prevents the
sensor from interacting with the environment, allowing its reading
to be controlled by the attacker. In other words, a high-frequency
audio signal at a specific frequency can bring these sensors into
a state of resonance, corrupting their outputs. The source of the
disruptive signal can be an external device situated next to the phone,
or even the phone\textquoteright s own speaker \citet{DBLP:conf/wisec/BlockNN17}.
Acoustic attacks on MEMS-based gyroscopes and accelerometers were
first presented by Son et al. in \citet{rockingdrones} in the context
of drones, and later shown by \citet{walnut,phonehome} to be applicable
to smartphone sensors as well. Tu et al. in \citet{DBLP:conf/uss/TuLLH18}
performed a comprehensive evaluation of out-of-band signal injection
methods to deliver adversarial control of embedded MEMS inertial sensors
on a wide variety of devices including self balancing scooters, stabilizers,
smartphones, VR headsets and other similar devices. Similarly, an
adversary equipped with a magnetic coil is able to spoof the outputs
of the magnetometer, an effect put to productive use in \citet{jiang}.
Recognizing the increasing risk caused by current and emerging sensor
spoofing attacks, the Industrial Control Systems Cyber Emergency Response
Team of the U.S. Department of Homeland Security (ICS-CERT) stated
recently that it considers orientation sensor attacks as a \textquotedbl threat
to critical infrastructure\textquotedbl\citet{homeland}.

Generally speaking, there are two types of spoofing attacks: corruption
attacks, which we refer to as sensor rocking attacks (following the
nomenclature of \citet{rockingdrones}) and rewriting attacks, which
we refer to as sensor rolling attacks (for reasons of symmetry). Sensor
rocking attacks replace the sensor readings with arbitrary corrupted
values which are unrelated to the external environment. For example,
the attacker can replace the sensor signal with a high frequency sine
wave or random noise. While the attacker cannot control the shape
of this corrupted signal, the attacker can turn the disruptive signal
on and off at will. In fact, \citet{phonehome} and \citet{jiang}
used this ability as a data transmission mechanism. Sensor rolling
attacks are a stronger class of attack, in which the attacker completely
replaces the sensor readings with values of their choosing. Since
the attacker can create any sensor readings including replaying previous
readings, defense methods that detect anomalies will not be effective
against rolling attacks. 

In this work, we replicate two types of acoustic attacks on the gyroscope,
as shown in \citet{phonehome} and \citet{DBLP:conf/uss/TuLLH18},
to collect data and test our defense methods. While \citet{phonehome}
used a piezoelectric speaker kept in close proximity to the phone,
\citet{DBLP:conf/uss/TuLLH18} used regular speakers connected to
an amplifier to attack the gyroscope from a distance. Both attacks
work by using the sensor's mechanical resonant frequency. To spoof
the magnetometer, we used a solenoid connected to a waveform generator
as magnetic field source similar to the methods of \citet{DBLP:conf/ches/ShoukryMTS13}.
The high sensitivity of the magnetometer makes it extremely vulnerable
to the presence of any external magnetic field, sometimes even to
the magnet in the phone's own speaker \citet{magnetometer}. 

\begin{table*}
\begin{centering}
\begin{tabular}{|c|c|c|}
\hline 
Device & Gyroscope & Magnetometer\tabularnewline
\hline 
\hline 
Samsung Galaxy S5 & InvenSense MPU-6500 & AKM AK09911c\tabularnewline
\hline 
Samsung Galaxy S6 & InvenSense MPU-6500 & Yamaha YAS532\tabularnewline
\hline 
LG Nexus 5X & Bosch Sensortec BMI160 & Bosch Sensortec BMM150\tabularnewline
\hline 
iPhone SE & InvenSense EMS-A & Alps Electric HSCDTD007\tabularnewline
\hline 
STM32L4 IoT Node & STMicroelectronics LSM6DSL & STMicroelectronics LIS3MDL\tabularnewline
\hline 
\end{tabular}
\par\end{centering}
\caption{\label{tab:Gyroscope-and-magnetometer}Gyroscope and magnetometer
sensors used in various test devices}
\end{table*}

\section{\label{sec:Defense-Methods}Defense Methods}

In this work we implement and evaluate two purely software-based approaches
for sensor spoofing detection. The first approach, SDI-1, uses machine
learning techniques applied to sensor output to detect anomalies.
The second approach, SDI-2, is a novel fusion-based detector which
works by examining multiple sensor outputs. Since these defenses apply
signal processing and machine learning, it is important to examine
the resource consumption of the defense methods both in terms of processing
time and of power consumption. It is also important to determine the
response time of the countermeasures. If the countermeasure has a
very high response time, it may be possible for an attacker to evade
detection by spoofing the outputs for just a very short amount of
time. To demonstrate the generic nature of our defenses across all
kinds of devices, we perform the attacks and test our defenses on
various smartphones, as well as an IoT node, as listed in Table \ref{tab:Gyroscope-and-magnetometer},
representing a wide variety of electronic devices with different constraints
in terms of CPU capabilities, memory and power consumption. 

\subsection{SDI-1: Machine Learning-Based Single Sensor Defense}

The key idea behind SDI-1 is to train a machine learning model that
can detect an anomaly (an attack) on the sensor output. To enable
this defense, the defender generates many traces of benign sensor
outputs and ideally traces of known attacks as well. Detection can
either be performed by a two-sided classifier, which is trained both
on benign and spoofed traces, or by a one-sided classifier, which
is only trained on benign traces; detection takes place when a new
trace deviates significantly from the benign traces. The advantage
of the single sensor approach is that it requires no additional inputs
other than the sensor readings themselves. Thus, it can be implemented
inside the sensor hardware (or inside its manufacturer-provided driver)
and does not require any high-level changes to the system. A possible
short-coming of this defense is that the two-sided classifier must
be trained on previously encountered and known attack traces. Any
new spoofing method which results in different attacker characteristics
will not be detected. This can be overcome by using a one-sided classifier
which only needs to be trained on benign traces. In this case, any
new trace which is significantly different from a benign trace will
be identified as an attack. The disadvantage of this approach is that
it only works for sensor rocking (corruption) attacks, and not for
sensor rolling (overwriting) attacks; indeed, if an adversary can
choose arbitrary values for the sensor, the attacker can simply replay
values recorded by the sensor in the past which cannot be identified
as anomalies.

Training a classifier directly on high-dimensional data, such as sensor
readings over time, is inefficient and can cause over-fitting. Thus,
before the learning algorithm operates on the traces, each trace must
be reduced into a small set of succinct features. In \citet{Das_CoRR,das_ndss}
the authors suggested a selection of features that are relevant for
positional sensor readings, and we use this set in our work as well. 

When designing our detector, we aimed to create a detector which is
both effective and explainable. Non-explainable classifiers, such
as ensemble-based methods or those based on deep learning, are less
appropriate in a fraud detection setting, since they do not clearly
indicate the reason for the detector\textquoteright s particular output.
We were interested in selecting a classifier that has a simple internal
structure and is therefore less sensitive to adversarial learning
scenarios, where the attacker has some access to the training set.
We looked for classifiers which had high accuracy and are less resource
intensive, so that our defense method can be applied on a wide range
of devices.

The single sensor defense can be implemented for all position sensors.
In this work, we focus on defenses against acoustic attacks targeting
the gyroscope and magnetic attacks targeting the magnetometer. We
briefly discuss defenses against acoustic attacks targeting the accelerometer
in Subsection \ref{subsec:Protecting-Against-Accelerometer}.

\subsection{\label{subsec:SDI-2:-Fusion-Based-Multiple}SDI-2: Fusion-Based Multiple
Sensor Defense}

The key insight behind the second defensive approach is that the defender
has an information advantage over the attacker whereby instead of
being limited to a single sensor, the defender can compare the current
readings of multiple different sensors measuring the same physical
phenomenon. If the sensors do not agree with each other, it can indicate
that an attack is in progress. The advantage of this approach is that
it works for both rocking and rolling attacks (i.e., even a completely
valid sensor trace replayed by the attacker will be detected if other
sensors on the system do not agree with it). Furthermore, this method
is generic and future-proof in the sense that it does not depend on
the characteristics of a specific attack method, but rather on the
immunity of the gyroscope to magnetic attacks and, correspondingly,
on the immunity of the magnetometer to acoustic attacks. To carry
out fusion-based defense in practice, we first derive the mathematical
relationships between the readings of different sensors, in this case
the gyroscope and the magnetometer. To this end, we apply some basic
Newtonian physics principles, as described below. Once the mathematical
relationships are identified, it is possible to use the waveform output
of one sensor to approximate the other sensor, or to use both sensors
to calculate the same intermediate waveform. Then, we can measure
the extent to which the two sensor readings agree, by applying some
sort of distance measure between the two waveforms.

Sensor fusion has its own advantages and disadvantages as a countermeasure,
as compared to single sensor detectors. Its main disadvantage is that
it has to accommodate at least twice the amount of measurement noise,
since it depends on multiple physical sensors. To highlight the different
between the methods, we first evaluate a threshold-based sensor fusion
detector based on a simple distance measure, namely the mean squared
error (MSE). We then show how this detector can be improved by combining
both sensor fusion and machine learning methods. 

The cornerstone of our fusion-based countermeasure is an equation
relating the readings of two different position sensors. The device's
sensor measurements are presented in a Cartesian coordinate system
$\left(X_{d},Y_{d},Z_{d}\right)$. This is the coordinate system (reference
frame) attached to the device. This coordinate system can be rotated
with respect to a fixed, Cartesian, or world coordinate system, $\left(X,Y,Z\right)$
, in which the axes follow the North-East-Down (NED) convention: $X=north$,
$Y=east$ and $Z=down$. The world frame is assumed to be inertial,
ignoring the rotational motion of the Earth. Note that the origins
of the two reference frames stay attached; translational degrees of
freedom are not accounted for. At some time instance t, the altitude
of the device frame with respect to the world frame is represented
by a set of time dependent Tait-Bryan angles $(\phi,\theta,\psi)$
. These are Euler angles where the sequence of rotations is x-y-z,
known also as roll, pitch and yaw. The transformation from the inertial
frame to the device frame is the rotation:

\begin{equation}
R(\phi,\theta,\psi)=R(\phi)R(\theta)R(\psi)\label{eq:Rotation-matrix-3D}
\end{equation}

where:
\begin{equation}
R(\psi)=\left[\begin{array}{ccc}
\cos\left(\psi\right) & \sin\left(\psi\right) & 0\\
-\sin\left(\psi\right) & \cos\left(\psi\right) & 0\\
0 & 0 & 1
\end{array}\right]\label{eq:Rotation-z}
\end{equation}
is a rotation around the initial $Z$ axis,
\begin{equation}
R(\theta)=\left[\begin{array}{ccc}
\cos\left(\theta\right) & 0 & -\sin\left(\theta\right)\\
0 & 1 & 0\\
\sin\left(\theta\right) & 0 & \cos\left(\theta\right)
\end{array}\right]\label{eq:Rotation-y}
\end{equation}
is a rotation around the intermediate $Y$ axis, and 
\begin{equation}
R(\phi)=\left[\begin{array}{ccc}
1 & 0 & 0\\
0 & \cos\left(\phi\right) & \sin\left(\phi\right)\\
0 & -\sin\left(\phi\right) & \cos\left(\phi\right)
\end{array}\right]\label{eq:Rotation-x}
\end{equation}
is a rotation around the final $X$ axis. A general vector is represented
in the rotated reference frame by:
\begin{equation}
\vec{G}_{d}=R(\phi,\theta,\psi)\vec{G}_{w}\label{eq:General-vector-transformation}
\end{equation}

To link the readings of the gyroscope and the magnetometer, we need
to express the angular velocity $\vec{\omega}$ in terms of the rotation
angles $\left(\psi,\theta,\phi\right)$. The angular velocity components
along the axes $\hat{\psi},\hat{\theta},\hat{\phi}$ perpendicular
to the rotations are given by:
\begin{equation}
\omega_{\psi}=\dot{\psi},\,\omega_{\theta}=\dot{\theta},\,\omega_{\phi}=\phi\label{eq:Angular-Velocity-angle-axes}
\end{equation}

The directions of these components of $\vec{\omega}$ cannot constitute
an orthogonal coordinate system; each rotation is made in a different
reference frame. The transformation matrices $R(\psi),R(\theta),R(\phi)$
can be used to project the angular velocity components on the Cartesian
coordinate system axes of the device frame \citet{ClassicalMechanicsGoldstein}.
We note that the first transformation is done by rotating around the
original z axis. Therefore, $\hat{\omega}_{\psi}$ is directed along
the original (world frame) z axis. In order to obtain the components
of $\vec{\omega}$ in the device frame we should use the full rotation
$R(\phi,\theta,\psi)$. The next rotation axis $\hat{\omega}_{\theta}$
coincides with the intermediate y axis and therefore should be transformed
by $R(\phi)$. The third axis of rotation $\hat{\omega}_{\theta}$
coincides with the final x axis and therefore does not undergo a transformation.
For each Cartesian component of $\vec{\omega}$ we can sum up the
contributions of the projections. As a result of this procedure, the
angular velocity in Cartesian coordinates of the device frame is given
by:
\begin{alignat}{1}
\omega_{xd}= & \dot{\phi}-\dot{\psi}\sin\left(\theta\right)\nonumber \\
\omega_{yd}= & \dot{\theta}\cos\left(\phi\right)+\dot{\psi}\cos\left(\theta\right)\sin\left(\phi\right)\nonumber \\
\omega_{zd}= & -\dot{\theta}\sin\left(\phi\right)+\dot{\psi}\cos\left(\theta\right)\cos\left(\phi\right)\label{eq:Angular-Velocity-Device-Frame}
\end{alignat}
We also note that the angular velocity transforms, as any other vector
would do, from the world frame to the device frame:

\begin{equation}
\vec{\omega}_{d}=R\left(\phi,\theta,\psi\right)\vec{\omega}_{w}\label{eq:world-frame-to-device-frame}
\end{equation}

In the remainder of this paper we exploit the fact that one can associate
the angles of rotation $(\phi,\theta,\psi)$ with the angular velocity
of the device, together with the equation \ref{eq:General-vector-transformation},
in order to relate the angular velocity in the device frame (measurements
of the gyroscope) to the rate of change in the magnetic field as measured
by the rotating device (measurements of the magnetometer).

\label{subsec:Magnetic-Field-time-deriv}\textbf{Magnetic Field Time
Derivative in the Device Frame.} Consider an arbitrary magnetic field,
constant and uniform in the world frame:
\[
\vec{B}_{w}=\left(B_{x},B_{y},B_{z}\right)
\]
thus:
\[
\frac{d\vec{B}_{w}}{dt}=0
\]
In order to obtain the magnetic field in the reference frame of the
device, the rotation matrix \ref{eq:Rotation-matrix-3D} is used:

\begin{equation}
\vec{B}_{d}(t)=R(t)\vec{B}_{w}\label{eq:B-in-device-frame-rotated}
\end{equation}
We take the derivative of \ref{eq:B-in-device-frame-rotated} with
respect to time to obtain: 
\[
\frac{d\vec{B}_{d}}{dt}=\frac{d\left(R(t)\vec{B}_{w}\right)}{dt}=\frac{d\left(R(t)\right)}{dt}\vec{B}_{w}
\]
We use the fact that the rotation matrix is orthogonal, $R^{-1}=R^{T}$,
and therefore $RR^{T}=R^{T}R=1$, and multiply the right-hand side
by $R^{T}(t)R(t)=1$:
\[
\frac{d\vec{B}_{d}}{dt}=\frac{d\left(R(t)\right)}{dt}R^{T}(t)R(t)\vec{B}_{w}
\]
or: 
\[
\frac{d\vec{B}_{d}}{dt}=\left[\frac{d\left(R(t)\right)}{dt}R^{T}(t)\right]\vec{B}_{d}.
\]
Calculations using \ref{eq:Angular-Velocity-Device-Frame} show that
$\frac{d\left(R(t)\right)}{dt}R^{T}(t)$ is a skew-symmetric matrix
obeying:
\[
\frac{d\left(R(t)\right)}{dt}R^{T}(t)=\left[\begin{array}{ccc}
0 & \omega_{zd} & -\omega_{yd}\\
-\omega_{zd} & 0 & \omega_{xd}\\
\omega_{yd} & -\omega_{xd} & 0
\end{array}\right]
\]
and multiplication of the matrix $\frac{d\left(R(t)\right)}{dt}R^{T}(t)$
with the magnetic field vector is equivalent to the negative of the
cross product of angular velocity with the magnetic field vector.
Thus, the final mathematical relationship between the measurements
of the magnetometer and the gyroscope is given by:
\begin{equation}
\frac{d\vec{B}_{d}}{dt}=-\vec{\omega}_{d}\times\vec{B}_{d}\label{eq:time-derivative-B-device-frame}
\end{equation}

A similar equation can be derived for the accelerometer-Doppler sensor
pair as well (see Section \ref{sec:Towards-Accelerometer-Doppler-Se}).

\begin{figure*}[t]
\includegraphics{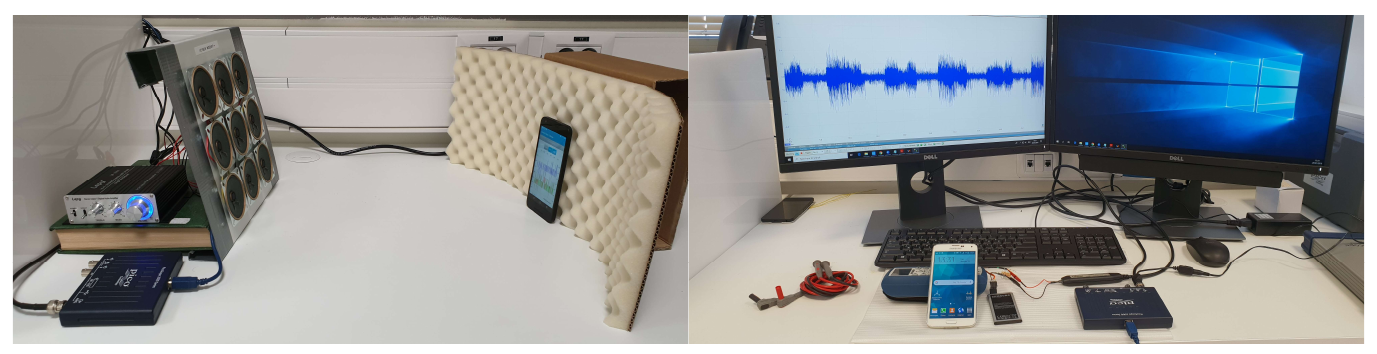}
\centering{}\caption{\label{fig:lab-setup}(a) Attacking the gyroscope (b) Carrying out
power analysis on Galaxy S5}
\end{figure*}

\section{\label{sec:Evaluation}Evaluation}

We evaluated the defenses for the gyroscope by first reproducing the
two acoustic attacks on the gyroscope as mentioned in \citet{phonehome}
and \citet{DBLP:conf/uss/TuLLH18}. To reproduce the attack of \citet{phonehome},
we used a PUI Audio APS2509S-T-R piezoelectric transducer connected
to a Picoscope 2206BMSO supported by Picoscope software v6.13.7.707
used as a waveform generator. To reproduce the attack in \citet{DBLP:conf/uss/TuLLH18},
we used a 4x2 dual channel PUI Audio AS06608PS-2-R speaker array with
8$\Omega$ impedance, connected to a Lepy LP-2051 audio amplifier
which received input from the same Picoscope 2206BMSO. To perform
automated frequency sweeps and frequency switches, we wrote a series
of Python script which could control the Picoscope using the libraries
provided by Picotech. 

To evaluate the defenses for the magnetometer, we used an air-core
solenoid with a 50$\varOmega$ impedance, connected to the same Picoscope
2206BMSO waveform generator through an amplifier. Our test devices,
as listed in Table \ref{tab:Gyroscope-and-magnetometer}, include
a variety of smartphones from multiple vendors, as well as an STM32L475VG
IoT node manufactured by STMicroelectronics. To collect the traces
from the phone, we wrote a custom Android application that timestamped
the sensor readings and uploaded them to an experiment server. The
server is capable of controlling various components like the frequency
of the wave, the number of traces to be collected and the duration
of each trace. The IoT node was running custom C++ code written using
the Mbed framework.

\subsection{\label{subsec:Methodology}Methodology}

The benign traces from all the phones were collected while the phone
was being subjected to typical user activities like walking, running,
at rest on the table, at rest in a pocket, and while the phone was
being randomly shaken (in which we also included the motions required
to play various video games). 

To carry out the acoustic attack, we first had to identify the resonance
frequency of the device. Tu et. al. in \citet{DBLP:conf/uss/TuLLH18}
listed the resonance frequency range of devices using the same sensor
model as our test devices. We used our experiment server to sweep
through the resonance frequency range and plotted the frequency against
the variance of the sensor reading to pin-point the resonance frequency
of the gyroscope. We determined the resonance frequency of the MPU
- 6500 series family of gyroscope used in out test smartphones to
be 27.243 kHz, and that of the LSM6DSL chip used in the IoT node to
be 19.718 kHz. 

Since \citet{phonehome} uses a piezoelectric speaker attached to
the phone, the attack traces were collected not only when the phone
was at rest but also when it was being moved (walking, running, shaking
etc.). However when the attack mentioned in \citet{DBLP:conf/uss/TuLLH18}
was carried out, the device was at rest as described in the paper.
The distance between the speaker array and the test device was 0.3
m. The IoT node was programmed to replicate the functioning of a self-balancing
scooter, one of the main test devices in \citet{DBLP:conf/uss/TuLLH18}.
A servo FS5103R motor was connected to the IoT node which rotated
based on the feedback of the gyroscope. The benign traces from the
IoT gyroscope included the sensor data when the device is at rest,
when subjected to a repetitive to and fro motion, and when subjected
to random shaking motions. 

To carry out the magnetic attack on the gyroscope, we found that we
were able to achieve maximum intensity and range of attack when the
frequency of the wave was 1 Hz. The magnetometer attack traces included
sensor traces when the solenoid was directed at the phone from different
directions and orientations; these variations affect different axes
differently. To simulate a rolling attack, in which the sensor reading
is arbitrarily determined by the attacker, we created random pairings
of benign gyroscope and magnetometer readings, each from a different,
independent measurement session.

In total, from each phone we had 500 benign traces (100 traces each
of walking, running, at rest on the table, at rest in a pocket, and
random shaking) of each sensor, 500 acoustic attack traces (250 traces
of each acoustic attack) of the gyroscope and 500 magnetic attack
traces of the magnetometer. From the IoT node we had 1500 benign traces
(500 traces each of at rest, under to and fro motion, and random shaking)
and 1500 acoustic attack traces (all traces collected under acoustic
attack mentioned in \citet{DBLP:conf/uss/TuLLH18}). The number of
benign and malicious traces collected were kept equal, to provide
balanced classes for the machine learning training algorithms. The
sensors were sampled at the highest possible sampling rate: 200 Hz
for the gyroscope and 100 Hz for the magnetometer. 

\subsection{SDI-1: Single Sensor Defense }

As mentioned earlier, training a classifier directly on high-dimensional
data, such as sensor readings over time, is inefficient and can cause
over-fitting. Thus, before the learning algorithm operates on the
traces, each trace must be reduced into a small set of succinct features.
Das et al. in \citet{das_ndss} identified a list of features relevant
for smartphone sensors in a different context. The data collected
from the gyroscope is a stream of timestamped real values. Since we
obtain the values from the three axes, the value is a vector consisting
of x, y and z values associated with a specific point in time. The
vector can be converted to a scalar by obtaining the L$_{2}$ norm
which is L$_{2}=\sqrt{x^{2}+y^{2}+z^{2}}$ . Another approach would
be to look at the readings of only one axis. Das et al. summaries
the characteristics of a sensor data stream by exploring a set of
25 features consisting of 10 temporal and 15 spectral features. With
the help of a domain expert we also identified a new feature to represent
the sensor data: max\_val\_fft, which is the maximum value of the
fast Fourier transform of the sensor data stream. To analyze the relative
importance of each feature, we used MATLAB\textquoteright s implementation
of the \textit{Relieff} algorithm \citet{DBLP:journals/apin/KononenkoSR97},
with k =20. The top ranking features and their corresponding weights
are listed in Table \ref{tab:Importance-of-each}.

\begin{table}[H]
\begin{centering}
\begin{tabular}{|>{\centering}p{1cm}|>{\centering}p{2cm}|>{\centering}p{3cm}|}
\hline 
Rank & Feature & Feature Importance Weight\tabularnewline
\hline 
\hline 
1 & Max val fft & 0.0520\tabularnewline
\hline 
2 & Max & 0.0514\tabularnewline
\hline 
3 & Mean & 0.0409\tabularnewline
\hline 
4 & Min & 0.0396\tabularnewline
\hline 
5 & Average Deviation & 0.0341\tabularnewline
\hline 
6 & RMS & 0.0329\tabularnewline
\hline 
7 & Standard Deviation & 0.0282\tabularnewline
\hline 
8 & ZCR & 0.0052\tabularnewline
\hline 
\end{tabular}
\par\end{centering}
\caption{\label{tab:Importance-of-each}Importance of each feature, according
to the \textit{Relieff} algorithm}
\end{table}

\subsubsection{Detecting Attack on Gyroscope on Smartphone}

After extracting the features from the raw traces, we used MATLAB's
Classification Learner tool to train and test various machine learning
models using a 10-fold cross validation scheme. The performance of
the various classifiers we evaluated is presented in Table \ref{tab:Accuracy-SDI1}.
As shown in the table, SDI-1 achieves a very high detection rate for
all of the devices we implemented. 

\begin{table*}
\centering{}%
\begin{tabular}{|c|c|c|c|c|c|}
\hline 
\multirow{1}{*}{Type} & \multirow{1}{*}{Classifier} & Galaxy S5 & Nexus 5X & Galaxy S6 & iPhone SE\tabularnewline
\hline 
\hline 
\multirow{3}{*}{Tree} & Simple & 98.9 & 92.9 & 96.7 & 100\tabularnewline
\cline{2-6} 
 & Medium & 98.9 & 96.2 & 96.7 & 100\tabularnewline
\cline{2-6} 
 & Complex & 98.9 & 96.2 & 96.7 & 100\tabularnewline
\hline 
Regression & Logistic Regression & 86.9 & 76.7 & 99.0 & 100\tabularnewline
\hline 
\multirow{5}{*}{SVM} & Linear & 85.7 & 73.8 & 98.6 & 100\tabularnewline
\cline{2-6} 
 & Quadratic & 97.0 & 91.0 & 99.0 & 100\tabularnewline
\cline{2-6} 
 & Cubic & 99.3 & 96.7 & 99.0 & 100\tabularnewline
\cline{2-6} 
 & Fine Gaussian & 99.6 & 97.6 & 99.5 & 99.7\tabularnewline
\cline{2-6} 
 & Medium Gaussian & 95.1 & 90.5 & 98.6 & 99.9\tabularnewline
\hline 
\multirow{6}{*}{KNN} & Fine & 99.0 & 96.2 & 100.0 & 99.9\tabularnewline
\cline{2-6} 
 & Coarse & 86.1 & 70.0 & 77.6 & 97.7\tabularnewline
\cline{2-6} 
 & Medium & 96.4 & 91.4 & 98.6 & 99.9\tabularnewline
\cline{2-6} 
 & Cosine & 94.4 & 92.9 & 97.6 & 99.9\tabularnewline
\cline{2-6} 
 & Cubic & 95.5 & 91.9 & 96.7 & 99.9\tabularnewline
\cline{2-6} 
 & Weighted & 97.4 & 96.7 & 99.5 & 99.9\tabularnewline
\hline 
\multirow{2}{*}{Ensemble} & Bagged Tree & 99.8 & 98.6 & 99.5 & 100\tabularnewline
\cline{2-6} 
 & Subspace KNN & 99.4 & 96.2 & 98.6 & 99.9\tabularnewline
\hline 
\end{tabular}\caption{\label{tab:Accuracy-SDI1}Offline accuracy (\%) of SDI-1 machine learning
classifiers for gyroscope using 10-fold cross validation}
\end{table*}

To evaluate the effectiveness of SDI-1 on the smartphone in an online
setting, we selected the classification tree algorithm due to its
consistently high accuracy and simple internal structure. We exported
the structure of the trained tree from MATLAB, and developed an app
in Android studio which implements the classification tree to detect
the attack on the phone. The app also made it possible to explore
different sampling window sizes, while keeping track of the true positives,
true negatives, false positives and false negatives so that we can
calculate the detection accuracy of the model. The app was initially
installed on a Galaxy S5. On initial testing we found that despite
the high accuracy shown when tested in MATLAB, our model had a very
high false positive and false negative rate, especially when the sampling
window was small when detecting in real-time. On inspecting the scatter
plot which plots the various features used by the classification tree,
we identified that the features we used were not able to separate
the attack and normal user activity. This indicated that the features
were not able to effectively separate between various acoustic attacks
and typical user activities in real-time.

To overcome this shortcoming, instead of extracting the features from
the L$_{2}$ norm, we extracted the features from the individual axes.
This required the calculation of eight features (Table \ref{tab:Importance-of-each})
on data from three axes. To reduce the number of calculations, we
decided to remove two features: ZCR (lowest rank) and max\_val\_fft
(calculation complexity). This leaves us with a total of six features
for each of the three axes, for a total of 18 features. The classification
tree was trained again using Classification Learner in MATLAB and
the model was implemented in the app. On testing this model showed
good performance, irrespective of the sampling window as shown in
Table \ref{tab:Real-time-accuracy}. To calculate the real time accuracies,
each attack detected when the phone was actually under was considered
as a true positive (TP) and each attack our defense failed to detect
was considered as a false negative (FN). During typical user activity
(no-attack) each falsely detected attack was considered a false positive
(FP) and rest as true negatives (TN). Accuracy was calculated using
the formula $Accuracy=\frac{TP+TN}{TP+TN+FP+FN}$ . Several minutes
of testing was done on all the devices under various attack and no-attack
scenarios to calculate its accuracy.

\begin{table}[H]
\begin{centering}
\begin{tabular}{|c|c|>{\centering}p{1.5cm}|c|}
\hline 
Phone & Sensor & Sampling Window (sec) & Accuracy (\%)\tabularnewline
\hline 
\hline 
\multirow{6}{*}{Galaxy S5} & \multirow{3}{*}{Gyroscope} & 1 & 98.42\tabularnewline
\cline{3-4} 
 &  & 2 & 98.18\tabularnewline
\cline{3-4} 
 &  & 5 & 98.33\tabularnewline
\cline{2-4} 
 & \multirow{3}{*}{Magnetometer} & 1 & 98.20\tabularnewline
\cline{3-4} 
 &  & 2 & 98.94\tabularnewline
\cline{3-4} 
 &  & 5 & 97.64\tabularnewline
\hline 
\multirow{6}{*}{Nexus 5X} & \multirow{3}{*}{Gyroscope} & 1 & 97.77\tabularnewline
\cline{3-4} 
 &  & 2 & 99.04\tabularnewline
\cline{3-4} 
 &  & 5 & 98.18\tabularnewline
\cline{2-4} 
 & \multirow{3}{*}{Magnetometer} & 1 & 99.08\tabularnewline
\cline{3-4} 
 &  & 2 & 98.75\tabularnewline
\cline{3-4} 
 &  & 5 & 98.33\tabularnewline
\hline 
\end{tabular}
\par\end{centering}
\caption{\label{tab:Real-time-accuracy}Real time accuracy (\%) of SDI-1 with
different sampling windows}
\end{table}

\textbf{One-Sided Classification:} As discussed earlier, in one-sided
classification the classifier is trained only using the benign data
and is tested on both the benign and malicious data. The main advantage
of this method is that, in contrast to a two-sided classification
which anticipate all attacks ahead of time, a one-sided classifier
will be effective against new attacks, as long as they sufficiently
deviate from the training data. 

The \textit{fitcsvm} function in MATLAB was used for one-sided classification,
based on the S5 data set. The data table consisting of the features
and the labels were divided into two parts (train and test). The benign
instances from the first part were used to generate the model using
the \textit{fitcsvm} function. The data from the second part was used
to test the model. The labels from the test table were removed and
the model was made to predict each instance as one or zero representing
an attack and no-attack respectively. The classifier gave 99.20\%
accuracy in this case.

We also tested one-sided classification of the Nexus 5X and the Galaxy
S6, albeit with smaller data sets, resulting in accuracies of 71.69\%
and 75.47\% for the Nexus 5X and Galaxy S6 respectively.

\subsubsection{Detecting Attack on Gyroscope on IoT node}

As mentioned in Section \ref{subsec:Methodology}, we collected 1500
benign and 1500 acoustic attack traces from the gyroscope. Considering
the severely constrained resources of the IoT node, we selected the
five simplest features from Table \ref{tab:Importance-of-each}: max,
mean, min, standard deviation and average deviation. The features
were extracted from the L$_{2}$ of the traces and then used to train
a simple tree using the Classification Learner tool of MATLAB, resulting
in a detection accuracy of 99.8\% in the offline model after 5-fold
cross validation. The tree which was trained using MATLAB was implemented
on the IoT node. We programmed an LED to turn on every time an attack
was detected. We also wrote a program to keep track of the true positives,
true negatives, false positives and false negatives. After extensive
testing under attack and under normal conditions, we obtained an accuracy
of 98.03\% with a sampling window of 5 ms. This proves that this defense
method is efficient and effective in a wide range of devices, even
under high resource constraints. In this case, unlike when using the
Galaxy S5, we were able to obtain high accuracy using the features
extracted from L$_{2}$ norms.

\subsubsection{Detecting Attack on Magnetometer on Smartphone}

Similar to the implementation of the single sensor defense on the
gyroscope, the single sensor defense was implemented on the magnetometer.
The accuracies of various machine learning models under K-fold cross
validation using the Classification Learner tool in MATLAB are shown
in Table \ref{tab:Accuracy-of-machine}. The trained classification
tree was implemented on the phone using our Android app. Unlike the
gyroscope, the model provided good accuracy (Table \ref{tab:Real-time-accuracy})
using features extracted from L$_{2}$norms. This shows that for \textit{Zero-order
sensors} like the magnetometer which measures the phone's static position
or orientation, features extracted from L$_{2}$ are sufficient to
differentiate between an attack and a no-attack scenario. However
for\textit{ First-order sensors }such as the gyroscope, which measures
the phone's rotation, features extracted from individual axes might
be better to differentiate between an attack and a no-attack scenario.
Single sensor defense for magnetometer was not carried out on the
IoT node. 

\begin{table}
\centering{}%
\begin{tabular}{|c|c|c|}
\hline 
\multirow{1}{*}{Type} & \multirow{1}{*}{Classifier} & \multicolumn{1}{c|}{Accuracy (\%)}\tabularnewline
\hline 
\multirow{3}{*}{Tree} & Simple & 91.4\tabularnewline
\cline{2-3} 
 & Medium & 90.3\tabularnewline
\cline{2-3} 
 & Complex & 80.4\tabularnewline
\hline 
Regression & Logistic Regression & 89.8\tabularnewline
\hline 
\multirow{5}{*}{SVM} & Linear & 87.4\tabularnewline
\cline{2-3} 
 & Quadratic & 90.4\tabularnewline
\cline{2-3} 
 & Cubic & 95.0\tabularnewline
\cline{2-3} 
 & Fine Gaussian & 93.8\tabularnewline
\cline{2-3} 
 & Medium Gaussian & 92.3\tabularnewline
\hline 
\multirow{6}{*}{KNN} & Fine & 93.1\tabularnewline
\cline{2-3} 
 & Medium & 91.2\tabularnewline
\cline{2-3} 
 & Coarse & 76.4\tabularnewline
\cline{2-3} 
 & Cosine & 91.2\tabularnewline
\cline{2-3} 
 & Cubic & 90.5\tabularnewline
\cline{2-3} 
 & Weighted & 93.0\tabularnewline
\hline 
\multirow{2}{*}{Ensemble} & Bagged Tree & 94.9\tabularnewline
\cline{2-3} 
 & Subspace KNN & 94.3\tabularnewline
\hline 
\end{tabular}\caption{\label{tab:Accuracy-of-machine}Offline accuracy (\%) of SDI-1 machine
learning classifiers for the magnetometer using 10-fold cross validation}
\end{table}

\subsection{SDI-2: Gyroscope-Magnetometer Sensor Fusion Defense}

In contrast to the machine learning defense presented in the previous
subsection, the sensor fusion countermeasure works by comparing the
output of the magnetometer, \textbf{$\vec{B}$}, to that of the gyroscope,
$\vec{\omega}$, as described in Subsection \ref{subsec:SDI-2:-Fusion-Based-Multiple}.
When the readings of the magnetometer and gyroscope are in agreement,
the equality 
\begin{equation}
-\vec{\omega}\times\vec{B}=\frac{d\vec{B}}{dt}\label{eq:gyro-mag-fusion-basic-equation}
\end{equation}
 should hold, regardless of the orientation of the phone. Therefore,
any difference between the two sides of Equation \ref{eq:gyro-mag-fusion-basic-equation}
should indicate that either the gyroscope or the magnetometer is being
spoofed. Translating this into practice, we first calculate the values
$\vec{\zeta}=-\vec{\omega}\times\vec{B}$ and $\vec{\eta}=\frac{d\vec{B}}{dt}$,
approximating $\frac{d\vec{B}}{dt}$ by the finite difference $\frac{\vec{B}\left(t\right)-\vec{B}\left(t-\Delta t\right)}{\Delta t}$.
All three components (x,y,z) of both $\vec{\zeta}$ and $\vec{\eta}$
are vectors of length $N$ for the given measurement period $T=N\Delta t$.
The mean square error (MSE) between the two signals is then given
by: 
\begin{equation}
MSE=\frac{1}{T}\sum_{i=1}^{T}\left(\left(\zeta_{i}^{x}-\eta_{i}^{x}\right)^{2}+\left(\zeta_{i}^{y}-\eta_{i}^{y}\right)^{2}+\left(\zeta_{i}^{z}-\eta_{i}^{z}\right)^{2}\right)\label{eq: MSE}
\end{equation}

It is important to note that the two sensors have different physical
characteristics. Specifically, the inexpensive Hall effect magnetometer
used on most phones has a slower response time, lower sensitivity,
and a higher noise level than the gyroscope. 

\begin{figure}
\begin{centering}
\includegraphics{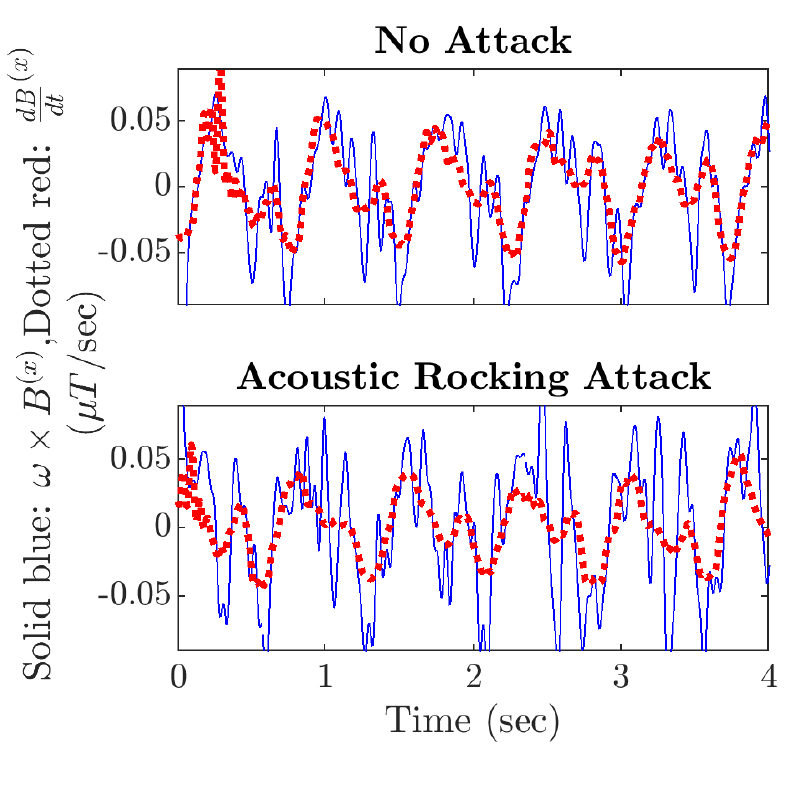}
\par\end{centering}
\caption{\label{fig:Sensor-fusion}A rocking attack can be detected by the
sensor fusion mechanism}
\end{figure}

Figure \ref{fig:Sensor-fusion} shows the output of a sensor fusion
calculation, captured on the Samsung Galaxy S5 phone, both under natural
conditions (top) and under a rocking attack (bottom). In both cases
the phone was placed in the researcher's pocket while the researcher
was walking around the lab. As seen in the figure, the values over
time of the x components of $-\vec{\omega}\times\vec{B}$ (solid blue)
and $\frac{d\vec{B}}{dt}$ (dotted red) are much closer on the top
half of the figure than on the bottom half. Nevertheless, the two
values plotted on the top graph are still not entirely identical,
due to the effects of the magnetometer's high measurement noise and
variations in external magnetic sources. As it is clear from the figure,
even when the device is not under attack, there is still a small difference
in the gyroscope and magnetometer reading. To mitigate these issues,
we need to specify a threshold value and assume that any deviations
below this threshold are normal. To identify the threshold, we calculated
the MSE between the two sensor signals under typical user activity,
under acoustic attack and under magnetic attack. Then, by using the
sensor fusion MSE as a single feature, we trained a single split binary
classification tree on MATLAB. Doing so we effectively instructed
MATLAB to create a threshold-based detector, choosing an ideal threshold.
When implementing the sensor fusion defense on the device, we can
use the same exact threshold which was identified by MATLAB. To implement
this method, we used a sampling window approach. An attack is identified
if within the sampling window, 80\% of the MSE's are above the threshold.
We implemented the sensor fusion on the Galaxy S5 using our Android
app. After extensive testing under normal conditions and in the face
of acoustic and magnetic adversary, the accuracy of sensor fusion
sampling is shown in Table \ref{tab:Real-time-accuracy-SDI2}. We
also carried out an offline threshold based sensor fusion defense
on an iPhone SE based on the data collected from its gyroscope and
magnetometer to obtain an accuracy of 74.4\%. 

\begin{table}
\begin{centering}
\begin{tabular}{|c|c|c|}
\hline 
Device & Sampling window (sec) & Accuracy (\%)\tabularnewline
\hline 
\hline 
\multirow{3}{*}{Salaxy S5} & 1 & 96.98\tabularnewline
\cline{2-3} 
 & 2  & 99.04\tabularnewline
\cline{2-3} 
 & 5  & 98.82\tabularnewline
\hline 
\multirow{3}{*}{Nexus 5X} & 1 & 98.68\tabularnewline
\cline{2-3} 
 & 2 & 98.38\tabularnewline
\cline{2-3} 
 & 5 & 97.95\tabularnewline
\hline 
\end{tabular}
\par\end{centering}
\caption{\label{tab:Real-time-accuracy-SDI2}Real time accuracy of of SDI-2
with different sampling windows}
\end{table}

\subsubsection{Sensor-Fusion on IoT node}

Similarly to the case of the smartphone, sensor fusion was also implemented
on the IoT node. Initially MSE was collected under normal conditions
and under attack conditions. A single split binary classification
tree using MSE as a single feature on MATLAB was used to identify
the threshold. Then, we applied the threshold-based sensor fusion
mechanism on the IoT node in real time. After testing under both normal
and attack (acoustic and magnetic) conditions, we obtained a detection
accuracy of 95.70\%. 

\subsubsection{Improving Sensor Fusion Using Machine Learning}

The advantage of the MSE threshold-based sensor fusion is its simple
structure. Once we identify the threshold, every MSE above the threshold
will be classified as an attack. Though our experiments on both the
Galaxy S5, Nexus 5X and the IoT node showed that sensor fusion is
highly effective as it is, it can be reinforced by using machine learning.
Similar to calculating the features from the L$_{2}$ norm in single
sensor defense, we can calculate the same set of features from the
MSE's within a sampling window. These MSE based features were used
to train various machine learning models using the Classification
Learner tool in MATLAB. The accuracies were calculated using different
schemes of K-fold cross validation. The accuracies of various machine
learning models trained and cross validated using Galaxy S5 data are
provided in Table \ref{tab:Offline-accuracy-of-SDi2}. As shown in
the table, these accuracies are equivalent to those of the threshold-only
defense, but we consider that this design may be more robust to intentional
disruption.

\begin{table}
\begin{centering}
\begin{tabular}{|c|c|c|}
\hline 
\multirow{2}{*}{Classifier} & \multicolumn{2}{c|}{Accuracy (\%)}\tabularnewline
\cline{2-3} 
 & 5-fold & 10-fold\tabularnewline
\hline 
\hline 
Fine tree & 97.4 & 97.2\tabularnewline
\hline 
Medium tree & 97.4 & 97.2\tabularnewline
\hline 
Quadratic SVM & 96.8 & 97.2\tabularnewline
\hline 
Fine KNN & 97.4 & 97.2\tabularnewline
\hline 
Bagged tree & 97.9 & 98.0\tabularnewline
\hline 
\end{tabular}
\par\end{centering}
\caption{\label{tab:Offline-accuracy-of-SDi2}Offline accuracy of SDI-2 using
multiple features extracted from MSE on Galaxy S5 traces}
\end{table}

\subsection{Real-Time Power Consumption and Performance Evaluation}

\begin{figure}[t]
\begin{centering}
\includegraphics[scale=0.9]{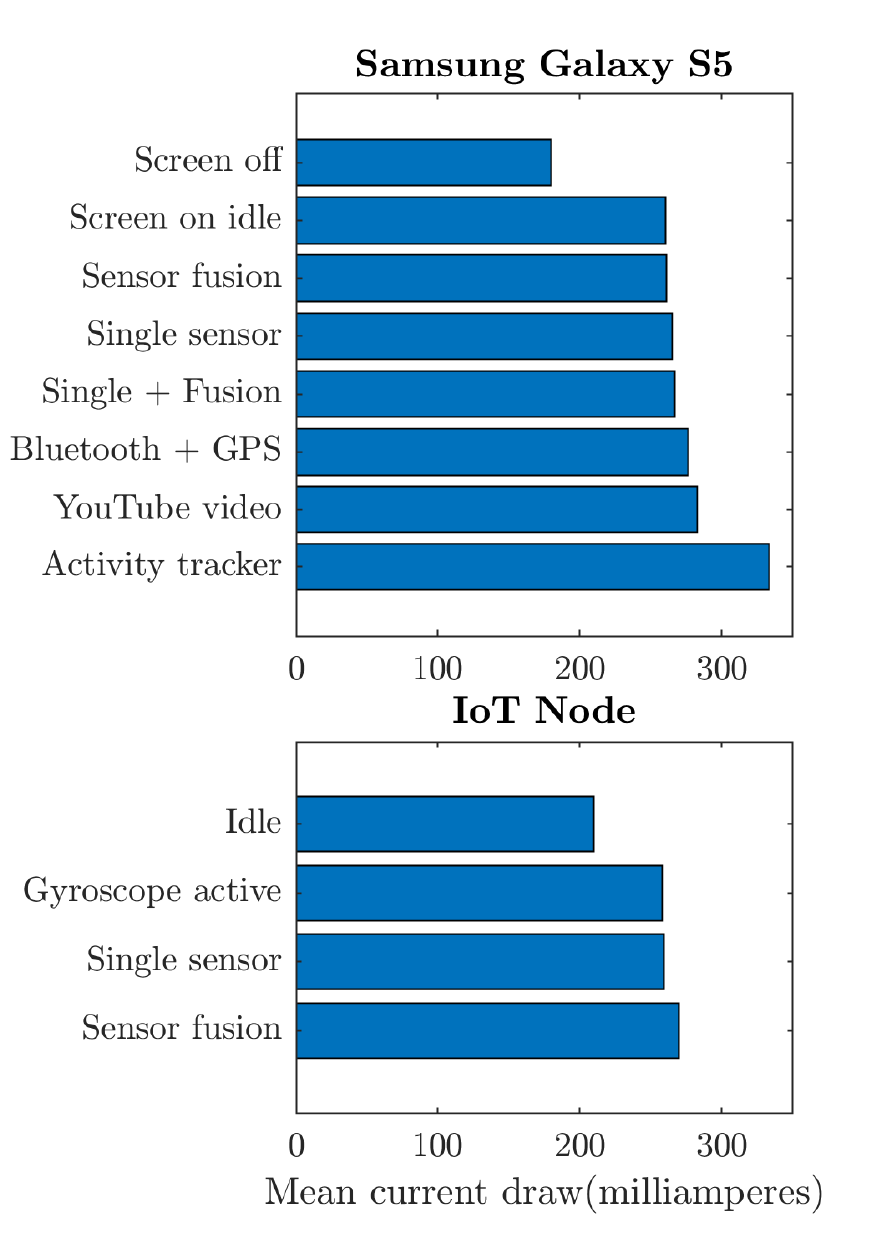}
\par\end{centering}
\caption{\label{fig:Power-consumption}The effect of enabling SDI on the instantaneous
power consumption of a smartphone (top) and an IoT node (bottom)}
\end{figure}

Since our targeted devices include smartphones and other low power
devices which have a limited energy reserve, any practical defense
must consume only a minimal amount of power. To show that our methods
provide this property, we measured their real-time power consumption
using an extrernal lab setup, as illustrated in Figure \ref{fig:lab-setup}.
To carry out the power analysis, we disconnected the battery from
the Galaxy S5 phone and routed it through a 0.2$\varOmega$ resistor
connected in parallel to a high-sensitivity Picotech TA046 800 MHz
Differential probe. The voltage drop on the probe was sampled and
stored on a PicoScope 2206BMSO oscilloscope. The traces were then
imported to MATLAB for analysis. As shown in Figure \ref{fig:Power-consumption},
our defenses consume a very small amount of power in excess to the
phones normal activities. To put matters in proportion, assuming that
the battery of the Galaxy S5 is at its full capacity of 2800 mAh and
that the phone is constantly turned on but left idle, a phone in which
our defense is always powered on will run out of battery 1.6 minutes
sooner than a phone without our defense, a difference hardly noticeable
by users.

Our generic fusion-based solution has very practical computational
requirements, making it feasible to implement on a variety of software
and hardware targets. Even though the mathematical model looks complex,
per sample calculation of the gryoscope-magnetometer fusion relationship
requires only a finite difference calculation (three subtractions),
a cross product (nine multiplications), a Euclidean distance calculation
(three multiplications), and finally a comparison to a threshold.
A performance evaluation was carried out on the Galaxy S5 running
Android version 5.0. Using Android Studio, we measured the real-time
CPU consumption of our countermeasure at a high resolution. Carrying
out the calculations for our defenses took between 70 and 150 microseconds.
Interestingly, this value was uncorrelated with the size of the sampling
window we selected, leading us to believe that most of this time was
actually spent on inter-process communications and UI updates, and
not on the calculation itself. The total detection time is the sum
of the sampling window size and the time taken for calculation. Which
means by using a sampling window size of 1 second , the time for detection
will be approximately 1.0001 seconds. When running at real-time at
a sensor sampling rate of 200 Hz (the highest rate possible on native
Android applications) our fusion calculations consumed only 0.5\%
of the phone\textquoteright s CPU. On our test device the app only
used 1.7\% of RAM, showing that the countermeasure is both effective
and feasible. We note that our application did not require special
user permissions nor modifications to the underlying operating system.
We believe that integrating our countermeasures into the device kernel
will cause its resource consumption to be even lower.

\section{\label{sec:Towards-Accelerometer-Doppler-Se}Towards Accelerometer-Doppler
Sensor Fusion}

There are spoofing attacks for the accelerometer which completely
control it (e.g. \citet{walnut}). Defending against this type of
attack requires that we correlate the accelerometer reading with another
sensor which measures linear motion. 

Doppler sensors are now being implemented in mobile phones, and they
will fortunately be capable of addressing this need. The Doppler sensor
measures the linear speed of the device, in meters per second (meter/sec),
by analyzing instantaneous shifts in the frequency of signals the
phone receives from Wi-Fi base stations and similar stationary radiation
sources, as a result of the Doppler effect. Similar to the gyroscope-magnetometer
sensor fusion, a similar mathematical relationship can be derived
for accelerometer-Doppler sensor fusion. Though this method is equally
effective, currently evaluating a full Doppler-based defense would
require us to make low-level modifications to the phone\textquoteright s
closed-source radio baseband stack. 

Before we can relate the signal received by the Doppler radar's antenna
to the measurements of the accelerometer, some processing is required.
Assume that the Doppler radar emits a \textbf{reference signal} in
the form:
\begin{alignat*}{1}
x\left(t\right)= & A_{0}\cos\left(2\pi f_{0}t\right)
\end{alignat*}
where $A_{0}$ is the amplitude of the transmitted signal. After this
signal is scattered by the moving phone, the Doppler radar sensor
receives a \textbf{surveillance signal} of the form:
\begin{alignat*}{1}
y\left(t\right)= & A_{R}\left(t\right)\cos\left(2\pi f_{0}t+\varphi\left(t\right)\right)+w\left(t\right)
\end{alignat*}
where $w(t)$ is some random noise, and $\varphi(t)$ is a time-dependent
phase:

\[
\varphi\left(t\right)=\frac{4\pi R\left(t\right)}{\lambda}=\frac{4\pi R\left(t\right)}{c}f_{0}
\]

$\lambda$ is the wavelength of the emitted EM wave, and$R\left(t\right)$
is the distance of the object from the Doppler radar, which obeys:
\[
R\left(t\right)=R_{0}+r\left(t\right)=R_{0}+\int_{0}^{t}v_{r}\left(t\right)dt.
\]
$R_{0}$ is the initial distance and $v_{r}$ is the radial velocity
of the object. The Doppler shift $\Delta f$ is proportional to the
time derivative of the phase $\varphi\left(t\right)$. The received
amplitude $A_{R}\left(t\right)$ depends on $R\left(t\right)$ and
the scattering cross section of the object.

Inside the Doppler radar, the received signal is converted to the
base-band Doppler shift signal. The conversion is performed by an
I/Q demodulator which mixes the received signal with the original
transmitted signal: 
\begin{alignat*}{1}
A_{R}\left(t\right)\cos\left(2\pi f_{0}t+\varphi\left(t\right)\right)\cos\left(2\pi f_{0}t\right)=
\end{alignat*}
\[
\frac{A_{R}\left(t\right)}{2}\cos\left(\varphi\left(t\right)\right)+\frac{A_{R}\left(t\right)}{2}\cos\left(4\pi f_{0}t+\varphi\left(t\right)\right)
\]
and with the $\frac{\pi}{2}$ phase shifted transmitted signal:

\begin{alignat*}{1}
A_{R}\left(t\right)\cos\left(2\pi f_{0}t+\varphi\left(t\right)\right)\sin\left(2\pi f_{0}t\right)=
\end{alignat*}

\[
-\frac{A_{R}\left(t\right)}{2}\sin\left(\varphi\left(t\right)\right)+\frac{A_{R}\left(t\right)}{2}\sin\left(4\pi f_{0}t+\varphi\left(t\right)\right)
\]
Next, a low-pass filter with a cutoff frequency of $f_{c}=500\ Hz$
is applied in order to remove the high frequency components which
yields the in-phase and quadrature components of the baseband signal,
plus some narrow-band noise:

\begin{alignat*}{1}
x_{I}\left(t\right)= & \frac{A_{R}\left(t\right)}{2}\cos\left(\varphi\left(t\right)\right)+\tilde{W}_{I}\left(t\right)
\end{alignat*}
\begin{alignat*}{1}
x_{Q}\left(t\right)= & -\frac{A_{R}\left(t\right)}{2}\sin\left(\varphi\left(t\right)\right)+\tilde{W}_{Q}\left(t\right)
\end{alignat*}
Note that the in-phase and quadrature components are the real and
imaginary parts of a base-band signal:
\begin{alignat*}{1}
Z_{BB}\left(t\right)= & \frac{A_{R}\left(t\right)}{2}\exp\left(-i\varphi\left(t\right)\right)+\tilde{W}_{BB}\left(t\right)
\end{alignat*}
We filter this signal again using a low-pass filter (in this case,
with a cutoff frequency of $f_{c}=100\ Hz$) in order to further reduce
noise. 

Phase relations between Ix and Qx indicate forward or backward movements.
Objects approaching the sensor $\left(\Delta f>0\right)$ generate
a -90 degree shift between Ix and Qx outputs. Objects moving away
from the sensor $\left(\Delta f<0\right)$ generate a 90 degree shift
between Ix and Qx outputs. 

In order to obtain the phase we use:
\begin{alignat*}{1}
\hat{\varphi}\left(t\right)= & \arctan\left(-\frac{x_{Q}\left(t\right)}{x_{I}\left(t\right)}\right)
\end{alignat*}
where $\hat{\varphi}$ includes the real phase and phase noise:
\begin{alignat*}{1}
\hat{\varphi}\left(t\right)= & \varphi\left(t\right)+n\left(t\right).
\end{alignat*}
We eliminate the constant (DC) part of the phase, which is proportional
to $R_{0}$, by using a high-pass filter ($f_{c}=0.5\ Hz$), and again
use a low-pass filter to eliminate noise and obtain the final result:
\begin{alignat}{1}
\tilde{\varphi}\left(t\right)= & \frac{4\pi}{c}f_{0}r\left(t\right)+\tilde{n}\left(t\right).\label{eq:Doppler-result-phase}
\end{alignat}
\textbf{Comparing the doppler radar to the accelerometer, }It is possible
to reconstruct the acceleration of the device from the Doppler shift
in the frequency of the electromagnetic (EM) wave emitted by a transmitter
(e.g., a Wi-Fi access point or cellular base station) and detected
by the device. In order to compare the measurements of the Doppler
radar to the measurements of the accelerometer one can either take
the derivative of Equation \ref{eq:Doppler-result-phase}, or, in
order to avoid the noise added in the process of differentiation,
or one can use the integral of the acceleration (measured by the accelerometer)
and apply a high-pass filter on the result to eliminate the constant
part of the integral. 

Note that this calculation requires the receiver to know the exact
shape of the transmitted waveform. In a classical Doppler radar setup
it is trivial to recover this waveform, since both the transmitter
and the receiver are in the same physical circuit. However, previous
work in the radar research community has shown that it is also possible
to recover the precise transmitted waveform of an external Wi-Fi receiver.

Assume that the transmitter and the receiver are moving in an instantaneous
relative velocity $\vec{v}\left(t\right)$, such that $\left|\vec{v}\left(t\right)\right|\ll c$,
where c is the speed of light, while the transmitter emits an EM wave
with frequency $f_{0}$. Without loss of generality, we assume that
the traveling wave has some wave vector $\vec{k}$ that forms an angle
$\theta\left(t\right)$ with the direction of the motion of the device.
Using these considerations, the Doppler shift is given by:

\begin{alignat*}{1}
\Delta f= & f'-f_{0}=\\
= & \frac{v\left(t\right)}{c}\cos\left(\theta\left(t\right)\right)f_{0}\\
= & \frac{v_{r}\left(t\right)}{c}f_{0}
\end{alignat*}
where $v=\left|\vec{v}\right|$, $v_{r}$ is the radial velocity (the
velocity in the direction of the line connecting the emitter and receiver)
and $f'$ is the shifted frequency. The sign of the radial velocity,
i.e. the velocity times the cosine of the angle between $\vec{v}$
and $\vec{k}$, indicates the sign of the shift. If the transmitter
and the receiver are moving towards each other, the sign of $v_{r}=v\cos\left(\theta\right)$
is positive and the detected EM wave is blueshifted. If, on the other
hand the transmitter and the receiver are moving away from each other,
the sign of $v_{r}=v\cos\left(\theta\right)$ is negative  and the
detected EM wave is redshifted. 

In order to simplify the math, in this paper we have only considered
cases in which the movement of the device is in the radial direction,
i.e.: $\vec{v}\parallel\vec{k}$; $\cos\left(\theta\right)=\pm1$,
and the acceleration is parallel to the velocity. In this specific
case we can derive a simplified form of the instantaneous acceleration,
in which the same convention about the direction of movement holds:
\[
a\left(t\right)=\frac{c}{f_{0}}\frac{d\left(\Delta f\right)}{dt}
\]

Note that by using the Doppler effect, one can only detect the relative
velocity in the direction of the line connecting the transmitter and
receiver. Reconstructing the acceleration in the case of an arbitrary
$\theta$ requires additional processing. One must also take into
account the gravitational acceleration added to the measurements of
the accelerometer; in this work we eliminated the effect of $\vec{g}$
by making sure that $\vec{g}$ stays perpendicular to the radial velocity
while measuring the acceleration of the device.

\section{\label{sec:Discussion}Discussion}

We presented two effective software-only methods for detecting acoustic
and magnetic attacks on the gyroscope and the magnetometer. We developed
and implemented our defenses, and performed detailed analysis on various
devices under various circumstances. One of the major advantages of
our defense methods is that they can be used for all kinds of devices.
Although the machine learning models require data collection and training,
this can be done externally, irrespective of the device, and only
the trained model need to be implemented on the device. In addition,
all of our defenses were independent of the size of the sampling window.
We were able to achieve good accuracy with a sampling window as small
as 5 ms on the IoT node, as mentioned in the previous section. Since
our defense method is purely software-based, implementing this on
existing devices requires just a simple firmware update and doesn't
require any expensive hardware modification. On inspection, we identified
that even the latest smartphones like the Galaxy S9 and One plus 5T
etc. use similar sensors to the ones used in our test devices. In
addition, not only the smartphones utilize these sensors -{}- a wide
range of electronic devices use these sensors to act as a bridge to
the outside physical world, which makes our work all the more important
in securing today's cyber physical systems. Our implementation on
the resource-constrained IoT node shows that even these resource-constrained
devices can be made safer against sensor spoofing attacks with no
additional hardware costs.

As we saw from the previous section, one of the main components determining
the accuracy is sufficient 'good' data. A reduction in the size of
the training data-set caused the reduction in accuracy when experimenting
with one-sided single sensor defense on the Nexus 5X and Galaxy S6.
Increasing the size of the data set used to train the model will have
a significant effect on the performance of the classifier. Manufacturers
who wish to implement these defenses can use a larger data-set, including
additional user activities from multiple users, to train the models
externally before implementing the defenses on the device. In addition,
more feature engineering can be done to create new features that can
better differentiate between an attack and a normal use. 

\subsection{Related Works}

Machine-learning based methods for detecting sensor malfunctions based
on a single sensor have already been considered in other domains,
such as the field of environmental sensor networks \citet{DBLP:journals/envsoft/HillM10}.
In \citet{DBLP:journals/envsoft/HillM10}, the authors demonstrated
the use of four data-driven methods for creating a one-step-ahead
prediction model to create a sensor anomaly detection system, based
on order q Markov models for different values of q. Even though this
method can fit many kinds of streaming data sets, it is not appropriate
for use in our scenario, where the characteristics of the signal can
change dramatically between consecutive samples even in benign situations.
In \citet{DBLP:conf/icmla/GunduzAD15} the authors reviewed a number
of proposed machine learning solutions pertaining to network layer
DoS attacks in wireless sensor networks. In \citet{DBLP:conf/uss/SikderAU17},
the authors proposed a context-aware intrusion detection system using
a machine learning based detection mechanism like Naive Bayes to detect
attacks which exploit the current insecure sensor management systems
of smart devices.

Sensor fusion was first discussed as a defense against corrupted sensor
readings by Chew et al. in \citet{DBLP:conf/srds/ChewM91}. In this
work, the authors presented a methodology for transforming a process
control program in a way that allows it to tolerate sensor failure.
In this methodology, a reliable abstract sensor is created by combining
information from several real sensors that measure the same physical
value. Based on this work, Ivanov et al. \citet{DBLP:journals/tecs/IvanovPL16}
discussed an optimal schedule for sampling from abstract sensors in
the presence of a spoofing adversary, and performed an experimental
validation of their methods on a simulation based on the Landshark
unmanned ground vehicle. In their work, Ivanov et al. sought to minimize
the intervals in which the system relies on sensor fusion by choosing
an optimal schedule in which the various sensors are sampled. While
this work discusses the best detection and counter detection strategies
for an abstract sensor, it does not implement a concrete sensor fusion
algorithm, as we present in our work. Delporte et al. made use of
positional sensor fusion in a constructive context in \citet{Delporte:2012aa}.
In this work, a world frame approximation of the gyroscope was obtained
while using a system equipped with only a magnetometer and an accelerometer.
Our system uses a simpler algorithm than that used by Delporte et
al. and makes fewer assumptions, since it is only interested in detecting
incongruities in the sensor reading and not in explicitly estimating
the sensor reading. In \citet{DBLP:journals/sensors/FanLL18} sensor
fusion algorithms were used for sensor bias estimations and adaptive
strategies. Nashimoto evaluated in detail the security of sensor fusion
by considering a sensor fusion scenario that involves measuring inclination,
with a combination of an accelerometer, gyroscope, and magnetometer
using Kalman filter in \citet{DBLP:conf/ccs/NashimotoSSS18}. In \citet{DBLP:conf/sp/KuneBCKRFKX13},
Kune et. al. used a software based method to mitigate EMI signal injection
attacks against analog sensors.

Shoukry et. al. in \citet{DBLP:conf/ccs/ShoukryMYDS15} developed
a physical challenge-response authentication scheme designed to protect
active sensing systems against physical attacks occurring in the analog
domain while Shin et. al. developed a method to bypass these timing
based sensor spoofing detection mechanism in \citet{DBLP:conf/woot/ShinSPKK16}.
Several countermeasures for positional sensor spoofing attacks were
presented by Trippel et al. in \citet{walnut}. These methods, including
in-phase sampling and randomized sampling, are mitigation countermeasures
which can degrade rolling attacks, transforming them into rocking
attacks, however, it is not clear how these countermeasures can be
used to detect an attack or generally gauge the confidence level of
a certain sensor measurement. An additional drawback of both of these
methods is that they require hardware modifications to the MEMS sensors.
These are highly integrated devices with relatively long development
cycles. They are also very specific defenses tailored to the WALNUT
attack, and they do not prevent against other methods of sensor spoofing.
Our proposed countermeasures only require changes to the phone software
and firmware, which is relatively quick to develop and deploy. Moreover,
our countermeasures make very few assumptions on the attacker\textquoteright s
method of attack, meaning that our countermeasure should serve well
against both present and future attacks.

As explained in previous sections, the attacks shown in \citet{phonehome}
and \citet{DBLP:conf/uss/TuLLH18} are the acoustic attacks we reproduced
to test our defenses. The list of features used by Das et. el. in
\citet{das_ndss} to develop sensor fingerprints served as the foundation
during the feature creation stage for our machine learning based single
sensor defense.

\subsection{\label{subsec:Protecting-Against-Accelerometer}Protecting Against
Attacks on Other Types of Sensors}

Our paper shows how to protect against attacks on the gyroscope and
the magnetometer. There are, however, also attacks on the accelerometer,
another common type of MEMS motion sensor which measures the linear
acceleration of the phone \citet{walnut}. To determine whether SDI
can protect against attacks against the accelerometer, we reproduced
an acoustic rocking attack on the accelerometer on the Galaxy S5 based
on \citet{walnut}, and deployed the SDI-1 defense against attack
using the same methodology we used to protect against gyroscope attacks.
Upon analyzing the data, we identified that the attack was very effective
and the sensor readings were clearly separable from the readings from
other user activities like running, walking, shaking etc. This led
to most of the machine learning classifiers having perfect accuracy
of 100\% in identifying an attack. However, there are spoofing attacks
for the accelerometer which completely control it (e.g. \citet{walnut}).
Defending against this type of attack requires the sensor fusion mechanism
of SDI-2, which means we must correlate the accelerometer reading
with another sensor which measures linear motion, as mentioned in
Section \ref{sec:Towards-Accelerometer-Doppler-Se}. 

\subsection{Is the Gyroscope Truly Invulnerable to Magnetic Attacks?}

\begin{figure*}[t]
\begin{centering}
\includegraphics{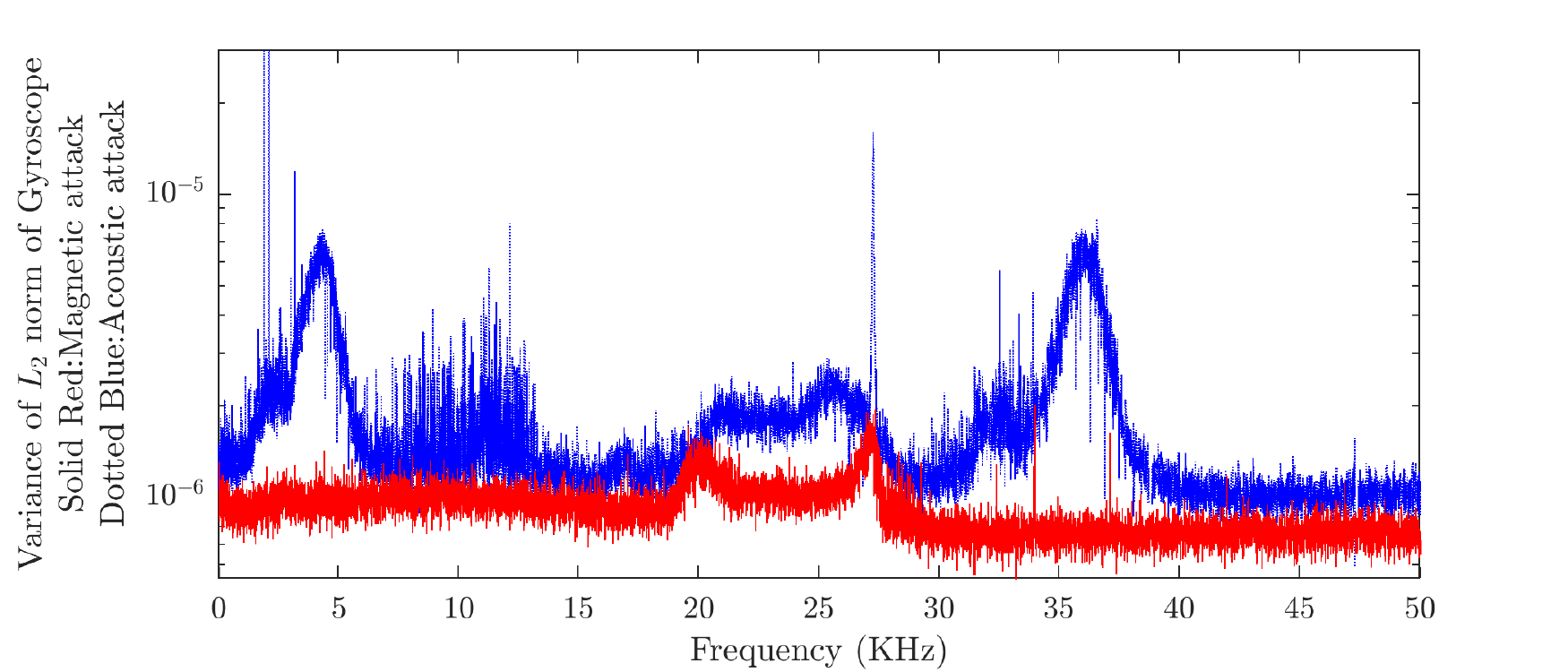}
\par\end{centering}
\caption{\label{fig:mag-attack-on-gyro}The effect of magnetic and acoustic
adversaries on the variance of a Samsung Galaxy S5 smartphone's gyroscope}
\end{figure*}

The initial phases of our research included identifying the effect
of magnetic and acoustic adversaries on the gyroscope and magnetometer
of the Galaxy S5. We did this by performing frequency sweeps using
the PicoScope with our experimental setup as explained in Section
\ref{sec:Evaluation}. The magnetometer was immune to an acoustic
adversary and vulnerable to magnetic adversary, as expected. Also,
as shown in many previous works, the gyroscope showed disturbance
in the face of an acoustic adversary. The gyroscope showed maximum
variance in its readings at its resonance frequency range at 27 kHz.
Interestingly, we identified that the gyroscope was showing disturbance
under a magnetic field as well. We were able to observe a spike in
the variance of the gyroscope readings under a magnetic adversary
which coincides exactly with the resonance frequency of the gyroscope
under the acoustic attack at 27 kHz as seen in fig \ref{fig:mag-attack-on-gyro}.
The magnitude of variance under a magnetic adversary is much smaller
than that of an acoustic adversary, but still significantly above
the noise level of a phone at rest. On inspecting the tear-down of
the device, we found that many of the important chips like the CPU,
RAM package, power management IC, gyroscope and accelerometer chip
etc are housed under a metallic covering. This metallic covering might
be causing the magnetic field to be converted to the corresponding
acoustic vibrations. The fact that the maximum variance under the
magnetic field coincides with the resonance frequency of the gyroscope
confirms this hypothesis. In our attempts, we were only able to use
the magnetic adversary as a rocking attack on the gyroscope. Unlike
the acoustic attack, due to its properties, the magnetic field is
difficult to direct and control as needed for a rolling attack. Generating
a directed magnetic field which can precisely control both the magnetometer
and the gyroscope can, in theory, cause our sensor fusion to fail.

\subsection{Responding to an Attack}

As explained in Section \ref{sec:Evaluation}, our defense can detect
attacks but are unable to prevent them. This leads to the natural
question: what should the phone do when there is major disagreement
between its various sensor readings?

To respond to an attack, we first have to identify which sensor has
been compromised. A device equipped with our single sensor defense
for both the gyroscope and magnetometer will be able to detect which
sensor is compromised, but will not be able to detect a rolling attack.
A system with the gyroscope-magnetometer sensor fusion defense will
be able to detect both rocking and rolling attacks, but will be unable
to identify the compromised sensor. An ideal system would have both
the single sensor and sensor fusion defenses implemented, allowing
it to detect both rocking and rolling attacks, and next to identify
the sensor that has been compromised.

Once we know which sensor has been compromised, one possible solution
is to attempt to simulate the corrupted sensor using the non-corrupted
one. While the performance of this simulated sensor will be degraded
compared to the original sensor (i.e. lower sensitivity, longer response
time, etc.), it will still be useful in many situations. In fact,
Delporte et al. \citet{Delporte:2012aa} were able to use only accelerometer
and magnetometer readings to create a \textquotedblleft virtual gyroscope\textquotedblright .
Another possible solution in the event of a sensor disagreement would
be to tweak the sensor readings until they both agree, effectively
halving the power of the attacker.

It seems that the optimal behavior in the case of sensor disagreement
which cannot be corrected would be to report an error condition to
the calling application, and leave the decision of how to respond
to the application developers. This will allow the application to
decide how to alert the user, and how to safely and intelligently
carry out at least parts of its original intended functionality, even
though it has low confidence in the readings of the sensor. How to
provide this degraded functionality in a usable and generic way remains
an open question.

\subsection{Improving Sensor Fusion}

In this work, we showed how to improve the reliability of one sensor
reading by comparing it to another sensor. We can generalize this
notion by comparing the sensor not just to other sensors, but to higher
order state indicators known to the phone. One such indicator that
might be combined with gyroscope readings in a sensor fusion algorithm
is the timing and location of touches on the phone\textquoteright s
touch screen. As shown in \citet{DBLP:conf/uss/CaiC11} and follow-up
works, the phone\textquoteright s position sensor readings are so
highly correlated with touches on the touchscreen that the gyroscope\textquoteright s
output alone can serve as a keylogger. We can reverse the direction
of inference, and consider what sort of gyroscope outputs should be
detected whenever a key is pressed. Incongruence could indicate that
the gyroscope is under a spoofing attack, or alternatively, that the
touch screen is under a touch injection attack \citet{DBLP:conf/woot/ShwartzCSO17}. 

In a wider sense, even higher-order notions, such as the activity
and general context of the phone, can be incorporated as inputs to
the sensor fusion algorithm. For example, when the screen\textquoteright s
display is off and its proximity sensor is active, one can reasonably
assume that the phone is in the user\textquoteright s pocket. As Unger
et al. have shown, \citet{DBLP:conf/huc/UngerRBGS14} data from the
phone\textquoteright s myriad sensors can determine many fine-grained
user contexts, such as periods when the user is eating, smoking, or
listening to music. Once the user context is established, the phone
can apply a sensor spoofing detection model fine-tuned to this context,
thereby achieving better performance. Also with more and more new
sensors being integrated into the devices (e.g., GPS, barometer, sonar,
lidar etc.), higher order sensor fusion has huge potential.

\subsection{Conclusion}

In this work, we developed, implemented and analyzed two new defenses
against acoustic and magnetic adversaries affecting the gyroscope
and magnetometer. Leveraging the information advantage the defender
has over the attacker, we applied sensor fusion methods to detect
when different sensor readings on the phone disagreed with each other. 

We showed how fusion-based defenses can be applied to the magnetometer
and gyroscope. Our software-only defense method can protect against
attacks which cannot be detected by other methods, including sensor
replay attacks (rolling attack). Sensor fusion defense can be augmented
by machine learning based single sensor defense methods. Most significantly,
our method has very realistic resource requirements and does not require
changes to the phone\textquoteright s hardware, drivers, or operating
system. Thus, it can be immediately put to use by phone manufacturers
as well as smartphone application developers. 

In future work, it would be interesting to flesh out the Doppler countermeasure,
especially as Doppler-equipped phones and 5G networks become more
prevalent. Future work could also focus on the evaluation of sensor
fusion defenses based on high-level context and touch events.

\bibliographystyle{elsarticle-harv}
\addcontentsline{toc}{section}{\refname}

\end{document}